\documentclass[aps,pre,twocolumn,groupedaddress,showpacs,showkeys,preprintnumbers]{revtex4}

\usepackage{graphicx}
\usepackage{color}
\usepackage[ansinew]{inputenc}
\usepackage{amsmath}
\usepackage{amsfonts}
\usepackage{amssymb}
\usepackage{dcolumn}

\usepackage{float}

\begin{document}
  \title {Morphology of Fine-Particle Monolayers Deposited on Nanopatterned Substrates}

  \author{N. A. M. Ara\'ujo}
  \affiliation{GCEP-Centro de F\'isica da Universidade do Minho, 4710-057 Braga, Portugal}
  \author{A. Cadilhe}
  \email[Electronic mail:\ ]{cadilhe@lanl.gov}
  \affiliation{T-12 Group, MS 268, Los Alamos National Laboratory, Los Alamos, NM 87545, USA}
  \affiliation{GCEP-Centro de F\'isica da Universidade do Minho, 4710-057 Braga, Portugal}
  \author{Vladimir Privman}
  \affiliation{Department of Physics, and Center for Advanced Materials Processing, Clarkson University, Potsdam, NY 13699, USA}
  \keywords{nanoparticle, patterned substrate, irreversible adsorption, jammed state\\{}\hphantom{\ }{}\\{}
{\large Posted as e-print 0709.3109 at www.arxiv.org}}
  %\date{$\mbox{September 18, 2007}$}

  \begin{abstract}
    We study the effect of the presence of a regular substrate pattern on the irreversible adsorption of nanosized and colloid particles. Deposition of disks  of radius $r_0$ is considered, with the allowed regions for their center attachment at the planar surface consisting of square cells arranged in a square lattice pattern.
    We study the jammed state properties of a generalized version of the random sequential adsorption model for different values of the cell size, $a$, and cell-cell separation, $b$.
       The model shows a surprisingly rich behavior in the space of the two dimensionless parameters $\alpha=a/2r_0$ and $\beta=b/2r_0$.
    Extensive Monte Carlo simulations for system sizes of $500\times500$ square lattice unit cells were performed by utilizing an efficient algorithm, to characterize the jammed state morphology.
  \end{abstract}

  \maketitle

  \section{Introduction}\label{Sec1}

    Monolayer and multilayer fine-particle deposits at surfaces are of interest for a wide range of applications, including photonic crystals, quantum dots, heterogeneous catalysts, sensors, and microarrays~\cite{Kumacheva02,Fustin03,Zhu04,Burda05,Lewis05,OConnor05,Yang05}.
    There has been a recent drive to explore and utilize particles smaller than the traditional colloid-size (few microns to sub-micron), namely down to nanoparticles (means dimensions of $0.01$ of a micron and smaller, i.e., sizes of order $10\,$nm and below).
    Quantification of the kinetics of synthesis, aggregation, and surface interactions of nanoparticles requires new experimental probes, but also new theoretical techniques.
    Furthermore, the surfaces with which fine particles interact, can now be pre-patterned to control and modify the particle attachment kinetics and the resulting deposit morphology~\cite{Jeon97,Chen02,Kumacheva02,Elimelech03,Cadilhe04,Lewis05,Joo06,Xia06}.
    Presently, experiments have produced patterns on the submicron scale, but the trend is down to nanoscale.
    Thus, deposition kinetics will be on an artificially formed lattice or another pattern, which can improve catalytic and reactivity properties of the particle-covered final surface.
    From a theoretical perspective, such processes pose interesting challenges, including identification of the parameters that control the properties of the resulting deposit.

    In the present work, we report a detailed study of the influence of a pattern consisting of square shaped cells in which centers of circular particles can land (e.g., projections of spherical fine particles depositing in a monolayer).
    The cells are in turn arranged on a planar substrate in a square lattice array.
    We consider the process of random sequential, fully irreversible adsorption of fixed size particles (disks).
    As usual for random sequential adsorption (RSA) processes~\cite{Bartelt91a,Bartelt91b,Bartelt91c,Bonnier01a,Bonnier01b,Evans93,Bartelt94,Bonnier94,Privman94,Brilliantov96,Privman00a,Privman00ed,Araujo06}, we assume that particles cannot overlap: the arriving disks, randomly and uniformly transported to the surface, are deposited only if they do not overlap earlier deposited ones (and provided their centers fall within the square pattern of the allowed-deposition cells).
    Deposition attempts of disks that do not satisfy these conditions are rejected, and the particles are assumed discarded (transported away from the surface).

    Thus the RSA model \cite{Evans93,Privman94,Brilliantov96,Privman00a,Privman00ed,Cadilhe07} assumes that the effects of the particle-particle interactions and particle-substrate interactions can be accounted for, approximately, by purely geometrical restrictions and features.
    The excluded volume constraint represents particle-particle interaction which is assumed short-range repulsion on length scales shorter than the particle sizes.
    Particle-substrate interaction is assumed to result in irreversible binding on the time scale of the process.
    Furthermore, the details of the particle transport to (and, for rejected particles, away from) the surface are lumped into the assumption of uniform flux of deposition attempts per unit surface area.
    In studies of RSA, one is interested in characterizing the jammed state morphology at large times, when a dense RSA deposit is formed and no available particle landing sites are left, as well as the approach to the jammed-state coverage. In this work we focus on the former aspect of the process: the jammed state properties.
    Despite its intrinsic simplicity, the RSA model provides a surprisingly rich set of limiting behaviors and morphologies \cite{Gonzalez74,Feder80a,Feder80b,Pomeau80,Swendsen81,Brosilow91,Privman91,Oliveira92,Evans93,Privman94,Lee97,Privman97,Privman00a,Privman00b,Loscar03,Cadilhe04,Cadilhe07,Subashiev07a,Subashiev07b}.

    The paper is organized as follows: In Section\ \ref{Sec2}, we start by defining the model, while in Section\ \ref{Sec3}, we analyze, analytically and numerically, the jammed state properties for the case when the excluded-volume interaction is functional only within individual landing cells. For geometries for which the excluded-volume interaction extends beyond individual landing cells, extensive Monte Carlo simulations are reported in Section\ \ref{Sec4} and \ref{Sec5}. We argue that the jamming coverage is not sufficient to study the jammed state, particularly in this case. The radial particle-particle correlation function is considered as the property of interest, in Section\ \ref{Sec5}, which also offers some concluding remarks.

  \section{The Model}\label{Sec2}

    Our primary goal is to study irreversible monolayer deposition of identical hard-core spherical particles on flat patterned substrates.
    This is obviously equivalent to deposition of disks with the hard-core, ``no overlap'' exclusion on a plane.
    The particle centers are only allowed to adsorb within well-defined bounded regions.
    These \textit{landing cells\/} will be for simplicity modelled as square ``landing areas'' allowed for the disks' centers.
    We point out that other cell shapes~\cite{Chen02,Childs02,Odom02,Paul03,Wu03,Cui04,Xia06} are feasible and relevant, e.g., rectangular, circular, etc.
    Moreover, we assume that the cells themselves are regularly positioned in a square lattice array, and, again, we note that other lattices are possible, e.g., the triangular lattice.
        The lattice structure has its own crystalographic \textit{unit cells}, which are larger than the landing cells.

    In the present work, we consider particles of fixed radius, $r_0$.
    This is obviously a theoretical idealization.
    Experimentally, the particle sizes and shapes will always have some dispersion.
    For dispersions above ${\cal O}(10\%)$ of the mean size, particles are considered polydispersed.
    However, syntheses of uniform spherical colloid and nanosize particles with polydispersity as low as $4\%$ has been reported~\cite{Guzelian96,Teranishi98,Teranishi99a,Teranishi99b,Peng00,Park01,Peng01a,Peng01b,Qu01,Toshima01,Battaglia02,Peng02,Yu02,Libert03a,Libert03b,Privman07}, so that the monodisperse approximation is quite realistic for many systems of interest.

    Thus, we assume that identical particles arrive with flux $F$, i.e., that the rate, per unit time, of deposition attempts of disk centers at the substrate is $F$ per unit area.
    The particles arrive to the surface randomly and uniformly.
      A deposition attempt fails if the disk's center falls outside the allowed-landing cells, or if the arriving disk overlaps one or more previously adsorbed ones, the latter mimicking the excluded volume interaction.
    These are the assumptions typical of the RSA model, and the binding is assumed irreversible: the particles do not detach from or move (diffuse) on the substrate on time scales of the dense deposit formation.
    This latter assumption is usually a very good approximation for colloid deposition~\cite{Privman00a,Privman00b}, but can be questioned, e.g., for protein deposition: indeed, studies of RSA-type models with particle rearrangement on the surface have also been reported~\cite{Privman93,Wang93a,Wang93b,Nielaba94,Grynberg95,Privman95b,Lavalle99,James00,Fusco01}.
        In the irreversible RSA model, the deposit density initially grows linearly with time, $t$.
    However, as the particle density increases, the hard-core exclusion leads to slow-down of the deposition process.
    Ultimately, for large times the ``jammed state'' is approached, at a density lower than that of close packing and with no long-range order in the particle positioning, but with no gaps left for any additional particle deposition.

    % #############################################################################
    \begin{figure}[t]
      \includegraphics[width=8.5cm]{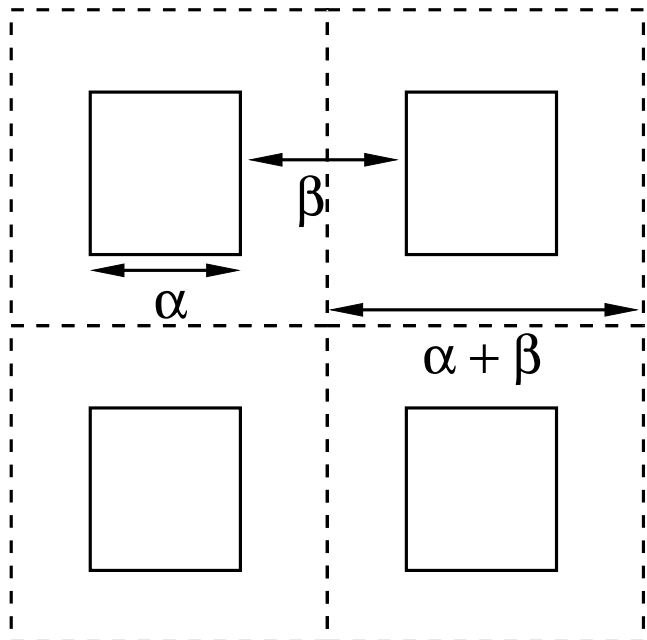}
      \caption
      {Dashed lines delineate the square lattice unit cells.
       Adsorption of disks can only take place when their centers fall inside the solid squares, i.e., the allowed-landing cells.
      \label{fig:pattern}}
    \end{figure}
    % #############################################################################

    For purposes of our modeling, we assume square landing-cells of linear size $a$, positioned in a square lattice with unit-cell size $a + b$ (obviously, $ a, b \geq 0$).
    This is, of course, an idealization, but we find that this geometry already offers a rich pattern of deposit morphologies.
We consider RSA of disks of radius $r_0$.
    Without loss of generality, we rescale the substrate lengths relative to the diameter of the disks.
    Specifically, we define

    \begin{equation}
      \alpha=\frac{a}{2r_0}\, ,\label{eq:alpha}
    \end{equation}

    \begin{equation}
      \beta=\frac{b}{2r_0}\, .\label{eq:beta}
    \end{equation}

    \noindent We comment that in studies of RSA the time is also usually rescaled, by a factor inversely proportional to the rate and particle ``volume,'' here $(\pi{}r_0^2F)^{-1}$.
    Figure~\ref{fig:pattern} illustrates the geometry of the lattice (in dimensionless units).

\begin{figure}[t]
      \includegraphics[width=8.5cm]{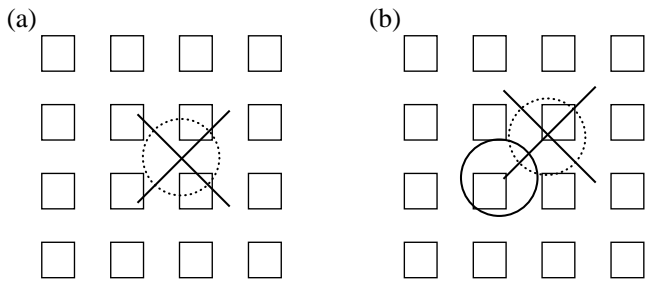}
      \caption
      {A disk fails adsorption onto the substrate because
        (a) its center does not fall within an allowed-landing cell, or
        (b) it overlaps with a previously adsorbed disk.
      \label{fig:model}
      }
    \end{figure}

    The deposition rules are illustrated in Fig.~\ref{fig:model}.
    Specifically, we do not allow particles to attach on top of each other, so that multilayer deposits are not formed~\cite{Onoda86,Adamczyk05}.
    Multilayer particle deposition was studied in various experimental and theoretical contexts~\cite{Bartelt91b,Privman92,Nielaba92,Nielaba95,Bardosova02}.
    In this regard our present model represents a generalized version of lattice RSA~\cite{Flory39,Renyi58,Bartelt91b,Evans93,Privman94,Privman00a,Privman00b,Talbot00,Weronski05,Adamczyk05,Cadilhe07}, on par with such generalizations as RSA of mixtures \cite{Talbot89a,Bartelt91c,Svrakic91,McLeod99,Bonnier01a,Hassan01,Hassan02,Cadilhe04,Araujo06,Subashiev07a,Subashiev07b,Loncarevic07} or deposition on finite-size substrates~\cite{Bartelt91a}.

    % #############################################################################
    % #############################################################################

  \section{The Jammed State}\label{Sec3}

    In the present model, the ``phase diagram'' is in the space of the two parameters, $(\alpha,\beta)$, the size of the cells and the distance between neighboring cells, in reduced units.
    The most obvious quantity to consider is the coverage, measured by the fraction of the total surface covered by particles, $\theta (t)$, in the jammed state for various values of $(\alpha, \beta)$.
    The jamming coverage, $\theta_J$, is a property of the state in which no further particles can adsorb, obtained in the large-time, $t$, limit,

      \begin{equation}
        \theta_J\equiv\lim_{t\rightarrow\infty}\theta(t)\,.
      \end{equation}

\begin{figure}[t]
      \includegraphics[width=8.5cm]{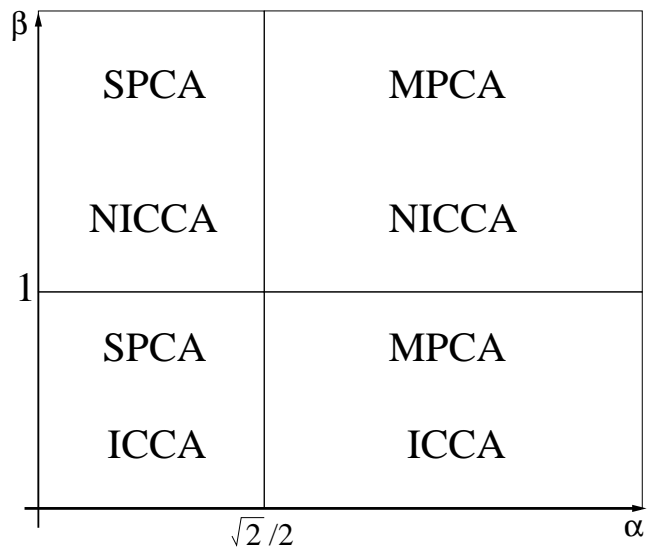}
      \caption
       {The major subdivisions in the two-parameter space.
        For cell-cell separation $\beta<1$ we have the \textit{interacting cell-cell adsorption\/} (ICCA), while for $\beta\ge{}1$ we have the \textit{non-interacting cell-cell adsorption\/} (NICCA).
        For cell sizes $\alpha<1/\sqrt{2}$ we have a \textit{single-particle-per-cell adsorption\/} (SPCA), while for $\alpha\ge{}1/\sqrt{2}$ we have \textit{multiparticle-per-cell adsorption} (MPCA).
        Limiting cases of the model, as well as subregions with interesting properties, are discussed in the text.
       }
       \label{fig:phase.diagram}
    \end{figure}

    If the landing cells are too far apart from their nearest neighbors, then particles at different cells will not be able to ``see'' each other through excluded volume.
    Specifically, for cells at a distance $\beta \ge 1$ apart from each other, particles (disks) attempting adsorption cannot overlap other disks in different cells.
    We denote this as the \textit{non-interacting cell-cell adsorption\/} (NICCA); see Fig.\ \ref{fig:phase.diagram}.
    For distances between cells $\beta < 1$ a particle attempting adsorption can overlap with a previously adsorbed one belonging to a different cell, thus leading to a failed deposition attempt.
    We denote this as the \textit{interacting cell-cell adsorption\/} (ICCA).

    % #############################################################################
    % #############################################################################

    One can choose the cell size to have a maximum, pre-determined close-packed number of particles inside it.
    For $\alpha<1/\sqrt{2}$, at most a single particle can adsorb inside any given cell.
    We denote this as \textit{single-particle-per-cell adsorption\/} (SPCA); see Fig~\ref{fig:phase.diagram}.
    For cells with $\alpha\ge{}1/\sqrt{2}$, more than a single particle can fit in the cell, and we denote this as \textit{multiparticle-per-cell adsorption\/} (MPCA).

    For the remainder of this section, we consider the NICCA case defined by $\beta\geq{}1$, which implies that a disk attempting adsorption with its center landing in a particular cell will never overlap a previously adsorbed disk in another cell.
      Thus, the global kinetics of deposition decouples into independent local kinetics at each landing cell.
      Therefore, for this range of $\beta$ values the model is equivalent to continuum RSA on finite-size substrates, with somewhat unusual boundary conditions that particles (disks) can ``stick out'' of the finite $\alpha \times \alpha$ square as long as their centers are within the square.
      The morphology and other physical quantities of interest of the global jammed state are determined by the jammed states for finite system sizes for the perscribed $\alpha$ value.

      % #############################################################################
      \begin{figure}[t]
        \includegraphics[width=8.5cm]{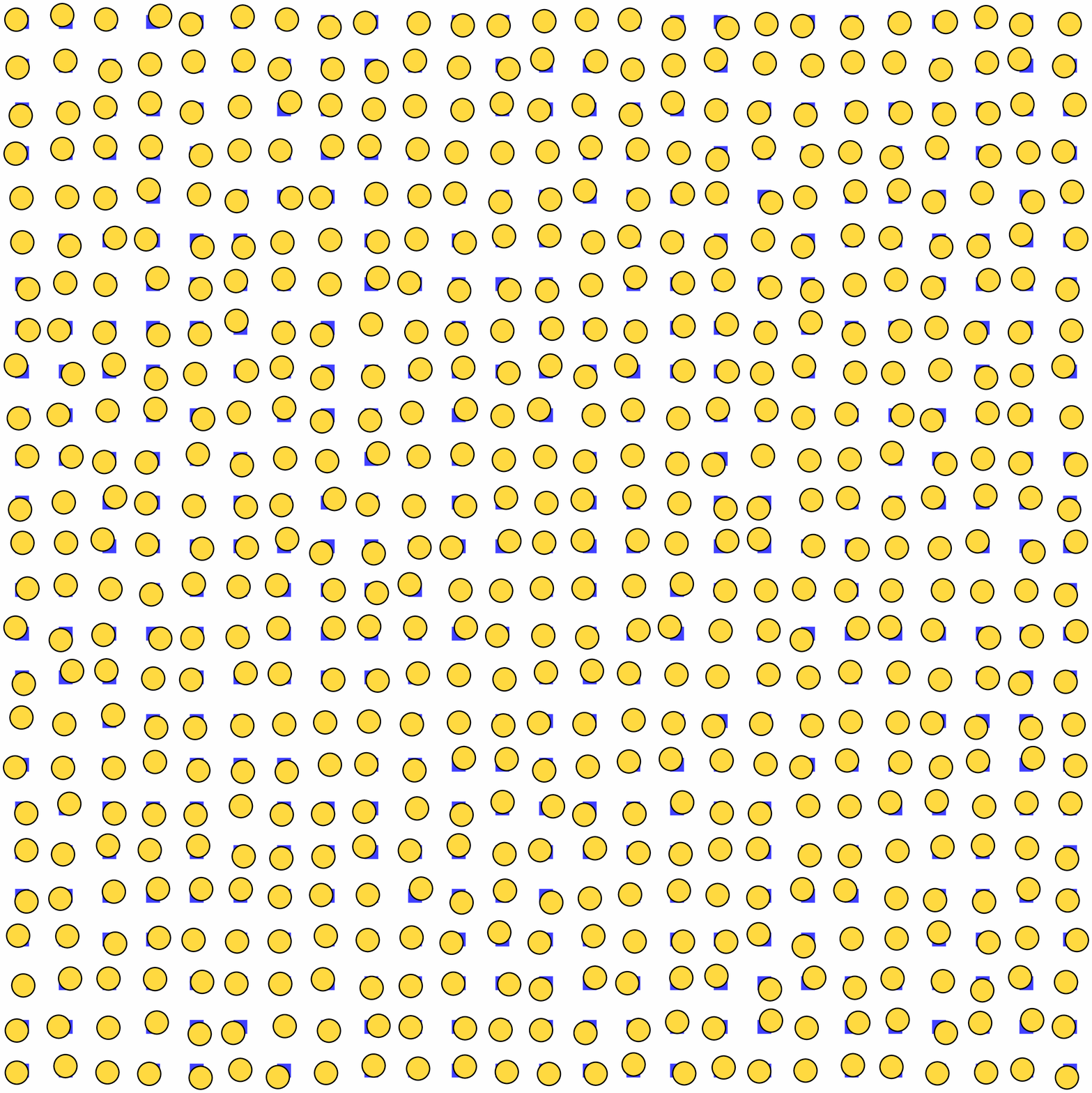}
        \caption
        {(Color online) Typical configuration of a region of $25\times25$ unit cells, for $\alpha= 0.6$ and 
        $\beta= 1.2$, in the jammed state.
        This snapshot corresponds to the NICCA-SPCA upper-left region in Fig.\ 
        \ref{fig:phase.diagram}.
        \label{fig:snapshotNICCASPCA}
        }
      \end{figure}
      % #############################################################################

      Let us consider the $\alpha$ and $\beta$ values to have NICCA with SPCA, where the latter case holds for $\alpha<1/\sqrt{2}$; see Fig.~\ref{fig:snapshotNICCASPCA} for a typical configuration.
      The kinetics corresponds to that of lattice RSA of monomers, since each cell is certain to have a single particle at the jammed state.
      The difference relative to the lattice RSA, is in the particle positions, which here are uncertain within the order of the size of the cell.
      While perhaps theoretically least interesting, such patterning provides for the most ``controlled'' particle deposition in applications.
      Since each cell ends up having a single particle, the jammed-state coverage is simply

      \begin{equation}
      \theta_{J}(\alpha, \beta)= \frac{\pi}{4(\alpha + \beta)^2}\,{},\label{equ:betage1alphalt1oversqrt2}
      \end{equation}

      \noindent which holds for $\beta\ge{}1$ and $0\le\alpha<1/\sqrt{2}$.
      Finally, for $\alpha= 0$ and $\beta\ge{}1$, we recover the true lattice RSA of monomers which are of unit-radius disks deposited at the centers of square cells $\beta \times \beta$.

      % #############################################################################
      \begin{figure}[t]
        \includegraphics[width=8.5cm]{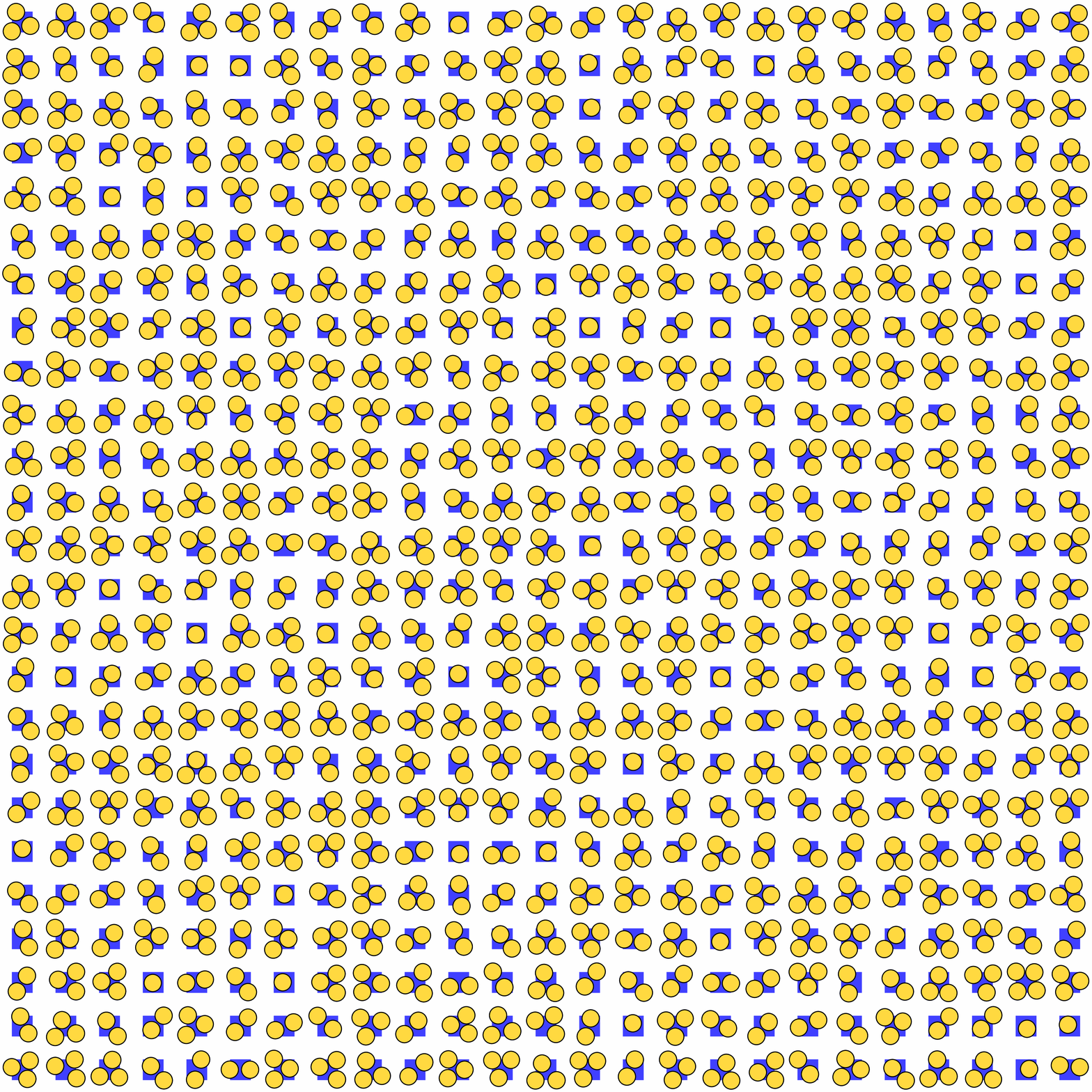}
        \caption
        {(Color online) Configuration of a region of $25\times25$ unit cells, for $\alpha= 1.2$ 
	and $\beta= 1.2$, in the jammed state.
         Particles attempting adsorption never overlap previously adsorbed ones in different cells, but 
	contrary to Fig.~\ref{fig:snapshotNICCASPCA} each cell can now adsorb more than one particle.
         Since $\beta=1.2>1$, the kinetics of adsorption at each cell is decoupled from that at other cells.
         \label{fig:snapshotNICCAMPCA}
        }
      \end{figure}
      % #############################################################################

      % #############################################################################
      \begin{figure}[t]
        \includegraphics[width=8.5cm]{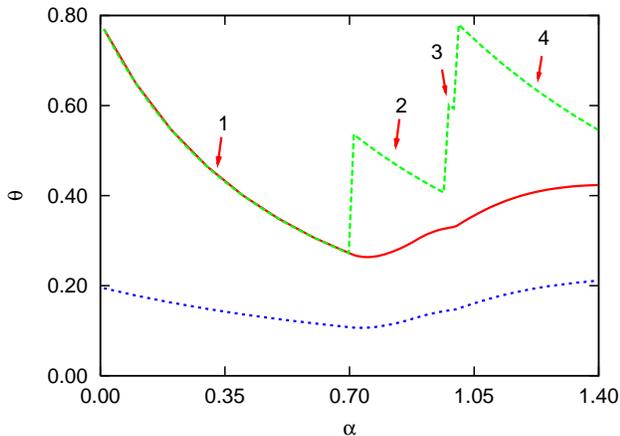}
        \caption
        {(Color online) Coverage values obtained by Monte Carlo simulation of our RSA model (solid line, the jammed-state coverage) and by direct calculation for the close-packed configurations (dashed line), both for $\beta= 1$.
        Notice the discontinuities in the values of the close-packed coverage as opposed to the smooth variation in the RSA case.
        The jammed-state coverages for $\beta= 2$, calculated according to relation (\ref{equ:thetaJbge1mapping}), are also shown for comparison (dotted line).
        \label{fig:coveragesNICCA}
        }
      \end{figure}
      % #############################################################################

      % #############################################################################
      \begin{table}[t]
        \caption
        {The table shows the jammed coverage values in the NICCA-MPCA case, i.e., for $\beta\ge{}1$ and $\alpha\ge{}1/\sqrt{2}$.
         In the first column, $n$ is the number of particles per unit cell of the close-packed situation.
         The values of $\alpha_n$, defined in the text, are shown in the second column.
         The third column gives the close-packed coverage values for the range of up-to-$n$ particles per cell.
         In the fourth column, the jamming coverage values, $\theta_J$, from simulations are presented.
         These values are for a representative choice $\beta = \beta^* = 1.2$ (see text), and for the $\alpha$ values shown in the second column.
         Finally, in the fifth column, the error estimate, $\sigma_J$, in the jamming coverages is given.
         \label{tab:coveragesbge1age1oversqrt2}
        }

        \begin{ruledtabular}
          \begin{tabular}{@{\quad}c@{\quad}c@{\quad}c@{\quad}c@{\quad}c@{\quad}}
            $n$&$\alpha_n$&$\theta_n$&$\theta_J(\alpha_n,\beta^*)$&$\sigma_{J}\times10^4$\\
            \hline\vspace*{-6pt}\\
            $2$&$\frac{1}{\sqrt{2}}$&$\frac{\pi}{2(\alpha + \beta)^2}$&$0.21594$&$0$\vspace*{3pt}\\
            $3$&$\frac{1}{2\sqrt{2}}(1 + \sqrt{3})$&$\frac{3\pi}{4(\alpha + \beta)^2}$&$0.26901$&$1.7$\vspace*{3pt}\\
            $4$&$1$&$\frac{\pi}{(\alpha + \beta)^2}$&$0.27471$&$1.7$\vspace*{3pt}\\
            $5$&$\sqrt{2}$&$\frac{5\pi}{4(\alpha + \beta)^2}$&$0.36148$&$1.0$\vspace*{3pt}\\
            $\infty$&$\infty$&$\frac{\pi}{2\sqrt{3}}$&$0.547063$&$0.75$\\
          \end{tabular}
        \end{ruledtabular}
      \end{table}
      % #############################################################################

      In the regime of NICCA with MPCA, a possibility opens up for having two or more adsorbed particles in each cell; see Fig.~\ref{fig:snapshotNICCAMPCA} for a typical configuration. Since the kinetics of deposition is \textit{decoupled\/} in the sense explained above, one can make some conclusions regarding the coverages for a given number of particles.

      Let us consider the case of up-to-two particles per cell as an example.
      It is obtained for values of $\alpha$ in the range $1/\sqrt{2}\leq \alpha < (1+\sqrt{3})/2\sqrt{2}$. The coverage for the close-packed (maximally packed) configurations is given by $\pi/2(\alpha + \beta)^2$ for two particles per cell and for $\beta\ge{}1$. The closed-packed coverage will change discontinuously at $\alpha = (1+\sqrt{3})/2\sqrt{2}$, as illustrated in Fig.~\ref{fig:coveragesNICCA}.
      However, our numerical results, also shown in Fig.~\ref{fig:coveragesNICCA}, indicate that the value of the jamming coverage for the RSA model, which for more than one particle per cell is less than the maximal coverage, remains continuous at $\alpha = (1+\sqrt{3})/2\sqrt{2}$ and also at $\alpha= 1/\sqrt{2}$. This behavior continues for larger number of particles per cell, see Fig.~\ref{fig:coveragesNICCA}. 
      In the close-packed problem, the highest coverage, at fixed $\beta$, occurs for the smallest cell, i.e., the smallest value of $\alpha = \alpha_n$ allowing the prescribed number of particles, $n$.
      Figure~\ref{fig:coveragesNICCA} illustrates the discontinuous coverage increments at $\alpha_n$, followed by a decrease $\propto (\alpha + \beta)^{-2}$.

      The simulated results for $\theta_J$, for $\alpha<1/\sqrt{2}$ follow the close-packed values, since the two problems coincide in this $\alpha$ range.
      However, at $\alpha= 1/\sqrt{2}$ the RSA coverage is continuous, since the probability of having a second adsorbed particle in any given cell remains small for $\alpha$ values above, but close to $\alpha_2 = 1/\sqrt{2}$. This property reflects the stochastic nature of the RSA model, i.e., in most cases the first particle adsorbs at a position inside the cell that blocks the chance for the second particle to adsorb later.  A similar observation applies as $\alpha$ crosses $\alpha_{n=3,4,\ldots}$.

      Our numerical results for $\theta_J$ were actually obtained for a representative $\beta$ value, $\beta^* = 1.2$.
      The coverage for any other $\beta\ge{}1$, for a given fixed value of $\alpha$, can then be calculated from

      \begin{equation}
        \theta_J(\alpha,\beta)= \left(\frac{\alpha + \beta^*}{\alpha + \beta}\right)^{\!\!\!2} \theta_J(\alpha,\beta^*)\,.
        \label{equ:thetaJbge1mapping}
      \end{equation}

      \noindent{}The error estimate for the coverage is given by

      \begin{equation}
        \sigma_J(\alpha,\beta)=\sqrt{\frac{\sum_{i=1}^N \theta_{Ji}^{\ 2}}{N}-\left(\frac{\sum_{i=1}^N 
        \theta_{Ji}}{N}\right)^{\!\!\!2}}\,,\label{equ:sigmabge1}
      \end{equation}

      \noindent where the index $i$ numbers the Monte Carlo runs, and $N$ stands for the total number of runs.
      From this definition, after some algebra one can show that

      \begin{equation}
        \sigma_J(\alpha,\beta)=\left(\frac{\alpha + \beta^*}{\alpha + \beta}\right)^{\!\!\!2} 
        \sigma_J(\alpha,\beta^*)\,.\label{equ:sigmabge1mapping}
      \end{equation}

      \noindent{}The above observation and the appropriate parameter and coverage values are summarized in Tables\ \ref{tab:coveragesbge1age1oversqrt2}, which in particular gives jamming coverage values $\theta_J (\alpha_n, \beta^*)$.

      % #############################################################################
      \begin{table}[t]
        \caption
        {Tabulation of numerically simulated RSA jammed-coverage values, at fixed $\beta= \beta^* = 1.2$, for the range of $\alpha$ from 0.72 to 1.40.
        The error estimates are also shown.
        \label{tab:NICCA}
        }
        \begin{ruledtabular}
          \begin{tabular}{@{\quad}c@{\qquad}c@{\qquad}c@{\qquad}c@{\quad}}
            $\alpha$&$\theta_J$&$\sigma_{\theta_J}\times10^4$\\
            \hline
            $0.72$&$0.21360$&$0.2$\\
            $0.74$&$0.21204$&$0.5$\\
            $0.75$&$0.21208$&$0.6$\\
            $0.76$&$0.21259$&$0.8$\\
            $0.78$&$0.21486$&$1.1$\\
            $0.80$&$0.21856$&$1.3$\\
            $0.82$&$0.22346$&$1.5$\\
            $0.84$&$0.22934$&$1.6$\\
            $0.85$&$0.23257$&$1.6$\\
            $0.86$&$0.23598$&$1.5$\\
            $0.88$&$0.24339$&$1.6$\\
            $0.90$&$0.25130$&$1.7$\\
            $0.92$&$0.25820$&$1.8$\\
            $0.94$&$0.26332$&$1.6$\\
            $0.95$&$0.26596$&$1.9$\\
            $0.96$&$0.26794$&$1.7$\\
            $0.97$&$0.26969$&$1.7$\\
            $0.98$&$0.27117$&$1.6$\\
            $0.99$&$0.27262$&$1.5$\\
            $1.02$&$0.28321$&$1.8$\\
            $1.04$&$0.29127$&$1.6$\\
            $1.05$&$0.29518$&$1.9$\\
            $1.06$&$0.29893$&$1.8$\\
            $1.08$&$0.30621$&$1.9$\\
            $1.10$&$0.31291$&$2.0$\\
            $1.12$&$0.31918$&$1.9$\\
            $1.14$&$0.32486$&$1.8$\\
            $1.16$&$0.33013$&$1.9$\\
            $1.18$&$0.33484$&$1.8$\\
            $1.20$&$0.33908$&$1.4$\\
            $1.22$&$0.34290$&$1.8$\\
            $1.24$&$0.34624$&$1.4$\\
            $1.26$&$0.34918$&$1.6$\\
            $1.28$&$0.35178$&$1.4$\\
            $1.32$&$0.35593$&$1.3$\\
            $1.34$&$0.35755$&$1.3$\\
            $1.36$&$0.35889$&$1.2$\\
            $1.38$&$0.36002$&$1.2$\\
            $1.40$&$0.36096$&$1.3$\\
          \end{tabular}
        \end{ruledtabular}
      \end{table}
      % #############################################################################

      The simulated jamming coverage values for varying $\alpha$ are given in Table\ \ref{tab:NICCA}. In the present RSA problem, for $\beta=\beta^*=1.2$ the minimum of the coverage is $\theta_{J_{\mbox{\tiny min}}}=0.21199 \pm 0.00006$, and it occurs for $\alpha_{\mbox{\tiny min}}= 0.745 \pm 0.005$.
      Finally, in the limit $\alpha\rightarrow\infty$ one recovers the well-known RSA of disks on continuum substrates \cite{Tanemura79,Feder80a,Hinrichsen86,Privman91,Meakin92,Evans93,Wang94,Privman00a,Privman00b,Cadilhe07}.
      Our value for the coverage is $0.54706 \pm 0.00008$, which should be compared to the range
$0.54700 \pm 0.00006$ recently estimated in \cite{Torquato06}. Note that the system size for which this new estimate of the jamming coverage was obtained was $2048\times2048$ particle diameters.
      The respective close-packed value is $\pi/2\sqrt{3}$, corresponding to the $n= \infty$ case in Table\ \ref{tab:coveragesbge1age1oversqrt2}.

  \section{Monte Carlo Simulations for Interacting Cell-Cell Adsorption}\label{Sec4}

    In this section, we begin our consideration of the ICCA regime. In this case, correlation can develop beyond single cells, and therefore extensive Monte Carlo simulations were warranted.
    Here we consider the jammed state coverage and morphology snapshots. However, particle-particle correlations have to be considered for a fuller quantitative description.
    This is taken up in the  next section.
    Counting system sizes as integer multiples of the unit cell side, $\alpha + \beta$, we took $500\times500$ as the finite system used in numerical simulations, unless otherwise stated.
    Periodic boundary conditions were applied both horizontally and vertically.

    A full description of the algorithm used, which represents an extension of the one introduced in \cite{Privman91,Wang94}, will be detailed in a separate publication \cite{Araujo07a}.
    Here, we summarize relevant details that might be pertinent to the direct interpretation of the results.
    We used the adaptive mesh-cell scheme \cite{Privman91,Wang94} to increase algorithmic efficiency, particularly at late times.
    Our adaptive mesh-cell scheme allowed for mesh-cell subdivision to half of the size, when the number of available mesh-cells dropped below $75\%$ of its original value.
    Such mesh-cell subdivisions allow to identify and exclude sub-cells already blocked by earlier deposited particles, thus improving the calculation speed of adsorption in the remaining ``open'' cells.
    One can show that this scheme does not bias the adsorption of particles \cite{Araujo07a}.
    We also comment that this approach can follow the time-dependent kinetics, though we only consider the jammed state in the present work.
    We also devised improved overlap tests of up to four surrounding (previously adsorbed) particles, which improved the algorithm performance.
    Another extension of the published algorithm \cite{Privman91,Wang94}, involved a natural implementation of the presence of the pattern, as detailed in \cite{Araujo07a}.

    The use of the algorithm allowed a detailed study of the structure of the jammed state.
    The number of realizations per simulation was $100$.
    We report the values of coverage and the corresponding error estimates in Tables~\ref{tab:coveragesbge1age1oversqrt2}, \ref{tab:NICCA}, and \ref{tab:ICCA}, for the various cases studied in the present work.

    Contrary to NICCA, in ICCA particles attempting adsorption at a particular cell, can overlap a particle previously adsorbed in a different cell.
    The ICCA regime is obtained for values of $\beta < 1$.
    The coverage values obtained in our simulations are given in Table~\ref{tab:ICCA}.
    As opposed to the NICCA case, in ICCA the kinetics of adsorption is no longer ``decoupled,'' and particles, or clumps of particles that belong to the same landing cell, not only follow the square positioning pattern of the landing cells but can also become correlated with particle in other cells. 
    The resulting deposit morphology and degree of ordering will depend on the geometrical parameters, as well as on the fully irreversible nature of the adsorption of the RSA model.
    In fact, depending on the values of $\alpha$ and $\beta$, particles attempting deposition can overlap others that belong to cells more distant than the nearest neighbor cells of the landing cell.
    Therefore, in addition to the jammed-state coverage, a more detailed study in the ICCA regime should also involve consideration of particle-particle correlations in the jammed state, as addressed in the next section.

    % #############################################################################
    \begin{table}[t]
      \caption
      {The table shows simulated values of the coverage, $\theta_J$, and the corresponding error estimates $\sigma_{J}$, for several values of $\alpha$ and $\beta$, in the ICCA-SPCA and ICCA-MPCA regions of the ``phase diagram'' shown in Fig.\ \ref{fig:phase.diagram}.
      \label{tab:ICCA}
      }
      \begin{ruledtabular}
        \begin{tabular}{@{\quad}c@{\qquad}c@{\qquad}c@{\qquad}c@{\quad}}
          $\alpha$&$\beta$&$\theta_J$&$\sigma_{\theta_J}\times10^4$\\
          \hline
          $0.1$&$0.14$&$0.5377\hphantom{0}$&$13$\\
          $0.2$&$0.2$&$0.50866$&$7.6$\\
          $0.2$&$0.28$&$0.54865$&$6.7$\\
          $0.2$&$0.5$&$0.46563$&$5.0$\\
          $0.2$&$0.8$&$0.54499$&$3.1$\\
          $0.3$&$0.42$&$0.48338$&$5.2$\\
          $0.4$&$0.2$&$0.52428$&$5.2$\\
          $0.4$&$0.56$&$0.54519$&$3.9$\\
          $0.5$&$0.7$&$0.54326$&$0.6$\\
          $0.6$&$0.2$&$0.53444$&$3.9$\\
          $0.8$&$0.2$&$0.55103$&$2.9$\\
          $0.8$&$0.8$&$0.33358$&$1.5$\\
          $1.0$&$0.2$&$0.53529$&$2.5$\\
          $1.0$&$0.8$&$0.39362$&$2.4$\\
          $1.2$&$0.2$&$0.53725$&$2.3$\\
          $1.2$&$0.8$&$0.45497$&$2.4$\\
        \end{tabular}
      \end{ruledtabular}
    \end{table}
    % #############################################################################

    Combining ICCA with SPCA, which holds for parameter values $\beta < 1$ and $\alpha < 1/\sqrt{2}$ and corresponds to the lower left region in Fig.~\ref{fig:phase.diagram}, can lead to particle configurations correlated beyond the cell pattern.
    We identify some limiting cases of interest.
    One such limit occurs for $\beta < 1$ and $\alpha \to 0$, which correspond to a well-defined square lattice structure for the deposition of disk centers.
    Since each adsorbed particle effectively ``shades'' a circle of unit radius, which is larger than the lattice constant $\beta<1$, deposition does not correspond to that of monomers as compared to the NICCA-SPCA case discussed above, as the adsorbed particle will surely block neighboring cells, and possibly more remote ones, depending on the value of $\beta$.
    However, taking $\beta= 0$, the system no longer has a pattern, regardless of the value of $\alpha$.
    Thus, the corner of the phase diagram near the point $(\alpha,\beta)= (0,0)$ is special, but we did not consider this region because it is more mathematically interesting than physically relevant: see recent literature on the kinetics of this type \cite{Bonnier01a,Hassan01}.

    % #############################################################################
    \begin{figure}[t]
      \includegraphics[width=8.5cm]{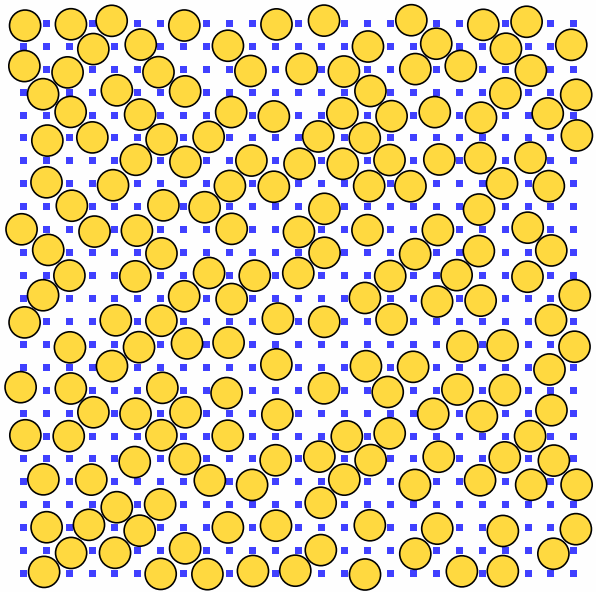}
      \caption
      {(Color online) Configuration of a region of $25\times25$ unit cells, for $\alpha= 0.2$ and $\beta= 0.5$, in the jammed state.
      A particle attempting adsorption can overlap a previously adsorbed particle in a different cell.
      This excluded volume interaction is responsible for correlations which result in locally-diagonal, semi-ordered domains as seen in this snapshot.
      \label{fig:snapshotICCASPCA}
      }
    \end{figure}
    % #############################################################################

      Due to the interaction between particles at different cells, during the deposition process correlations can develop, resulting in nontrivial local particle arrangements.
      These correlations stem from the excluded volume interaction between particles upon adsorption and they can induce local semi-crystalline order.
      Visually, see Fig.\ \ref{fig:snapshotICCASPCA}, the ``crystallites'' are oriented along the diagonal direction of the square lattice of square cells.
      However, it is well known \cite{Feder80a,Pomeau80,Swendsen81} that the RSA process alone cannot impose long-range ordering.
      Indeed, the particle correlations in RSA are usually rather short-range.
      We find that the patterning of the surface can induce ordering over several lattice spacings, which reflects the symmetry of the underlying pattern.
      More generally, the ordering should also depend on the shape of the deposited objects \cite{Brosilow91,Privman92}.
      Thus, the pattern does influence the creation of ordered structures in an otherwise uniform deposition process.
      However, the stochastic RSA dynamics tends to prevent the long-range order of the pattern from being fully ``imprinted'' in the deposited particle configuration, as observed in Fig.\ \ref{fig:snapshotICCASPCA}.

      % #############################################################################
      \begin{figure}[t]
        \includegraphics[width=8.5cm]{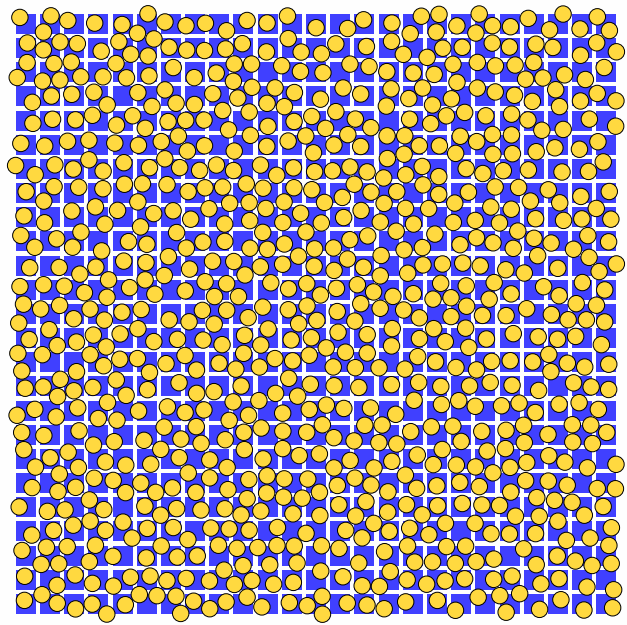}
        \caption
        {(Color online) Configuration of a region of $25\times25$ unit cells, for $\alpha= 1.2$ and $\beta= 0.2$, in the jammed state.
         For this low $\beta$ value, the probability of a particle attempting adsorption to overlap with one in a neighboring cell is appreciable, thus building up a somewhat longer-range diagonal semi-ordering than that seen for the parameter values of Fig.~\ref{fig:snapshotICCASPCA}.
         \label{fig:snapshotICCAMPCA}
        }
      \end{figure}
      % #############################################################################

      Combining ICCA with MPCA, with the parameter values $\beta < 1$ and $\alpha \ge 1/\sqrt{2}$, corresponds to the lower right region of the ``phase diagram'' in Fig.~\ref{fig:phase.diagram}. In this regime, the excluded volume interaction can lead to deposit morphology with semi-ordering beyond a single cell. However, the overlap with particles in neighboring cells can extend at most up to the second-nearest-neighbor cells (diagonally neighboring cells). In fact, for $1>\beta\ge{}1/\sqrt{2}$ (with $\alpha \ge 1/\sqrt{2}$), the overlap can be at most up to the nearest-neighbor cells.
      As $\alpha$ increases, one expects a lower impact of the cell-cell excluded volume interaction on the morphology.
      Consequently, for more cell-cell exclusion effects, we need smaller cell sizes, with limited number of particles (but at least two) allowed in each.
      The cell-cell exclusion leads to a further reduction of the average cell population, is illustrated in Fig.~\ref{fig:snapshotICCAMPCA}, where $\alpha= 1.2$ and $\beta= 0.2$.
      For these values of the parameters, each cell have enough area to accommodate up to four particles, but excluded volume interaction due to nearest-neighbor cells for this low value of $\beta$ substantially lowers the average cell population, as compared, e.g., with that of Fig.~\ref{fig:snapshotNICCAMPCA} for $\alpha= 1.2$ and $\beta= 1.2$: the average number of particles per cell in the case of $\alpha= 1.2$ and $\beta= 1.2$ is $2.487\pm 0.001$, while that of $\alpha=1.2$ and $\beta= 0.2$ is $1.3407\pm 0.0006$.
      However, the case of $\beta= 0.2$ still has a higher value of the coverage, $0.5373 \pm 0.0002$, while for $\beta= 1.2$ the value of the coverage is lower, $0.3391 \pm 0.0001$, because of more void space between the cells.

  \section{Interparticle Distribution Function}\label{Sec5}

    To further characterize the jammed state, we introduce the distribution function of the distances, $r$, between the centers of the adsorbed particles, $P_{\mbox{rad}}(\alpha, \beta; r)$.  
    The system is translationally invariant in terms of the integer multiples of the unit cell size ($\alpha + \beta$), because we use periodic boundary conditions both horizontally and vertically.
    Therefore, only displacement vectors between particle centers matter when studying particle-particle correlations.
    For convenience, in order to avoid discussion of the ``connected'' part vs.\ the full correlation function, we normalized the correlations by counting only distances between pairs of particles within a cutoff distance, $R$.
    The distances $r$ and $R$ will be assumed dimensionless, measured in units of the particle diameter.
    We found it appropriate to limit our study to separations up to $R=5 (\alpha + \beta)$.
    Since our results were not accurate enough to study possible weak singularities that can develop in the jammed-state RSA correlations at particle-particle contact \cite{Feder80a,Pomeau80,Swendsen81}, at $r=1$, and we were interested in the tendency for semi-ordering on length scales of several unit cells, we found it convenient to define

    \begin{widetext}
      \begin{equation}
        P_{\mbox{rad}}(\alpha, \beta; r)=
        \frac{\mbox{\ Number of pairs of particles with distances in\ }(r,r+\mbox{d}r)\mbox{\ }}
        {\ {}r{\,}\mbox{d}r{\,}(\mbox{Total number of pairs of particles at distances }<R)\ \!{}}\,.
        \label{equ:RDF_definition}
      \end{equation}
    \end{widetext}

    \noindent This distribution function is normalized as follows,

    \begin{equation}
      \int_0^R P_{\mbox{rad}}(\alpha, \beta; r)\ \!r\,\mbox{d}r= 1\,.
      \label{equ:RDF_normalization}
    \end{equation}

    The shape of the distribution function in the jammed state depends on the values of $\alpha$ and $\beta$.
    The position of the first peak measures typical distances between the closest particles.
    To better understand the role of $\alpha$ and $\beta$, we considered three families of distribution functions, as defined in the following subsections.
    The time-dependent kinetics of a distribution function will be studied in a forthcoming article \cite{Araujo07a}.

    % #############################################################################
    \begin{figure}[t]
      \includegraphics[width=8.5cm]{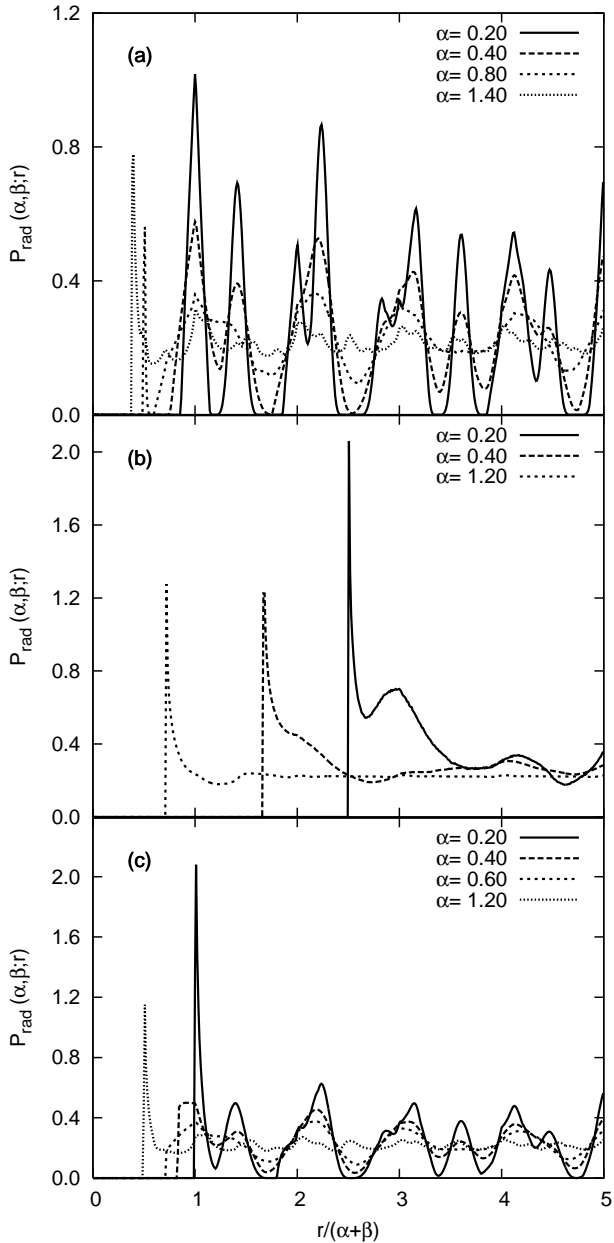}
      \caption
      {Plots of the distribution function for various values of $\alpha$, at fixed values of $\beta$:
       (a) $\beta= 1.2$, with $\alpha= 0.2$, $0.4$, $0.8$, and $1.4$; (b) $\beta= 0.2$, with $\alpha=0.2$, $0.4$, and $1.2$;
       (c) $\beta= 0.8$, with $\alpha= 0.2$, $0.4$, $0.6$, and $1.2$.
       \label{fig:RDFFixedBeta}
      }
    \end{figure}
    % #############################################################################

    \subsection{Effect of varying $\alpha$ on the distribution function}

      We start by studying the influence of varying $\alpha$ on the semi-ordering of the jammed state.
      We carried out a series of simulations at fixed $\beta=1.2$ and varied $\alpha$.
      The results are shown in Fig.\ \ref{fig:RDFFixedBeta}. The dimensionless center-center distance, $r\,$---$\,$the horizontal axis of the plot$\,$---$\,$was further rescaled in terms of the unit cell size, to $r/(\alpha + \beta )$, see Fig.\ \ref{fig:RDFFixedBeta}.
      Keeping $\beta\ge 1$, here $\beta=1.2$, corresponds to the NICCA case.
      In the NICCA-SPCA case, the first peak appears at a unit (rescaled) distance, since the distance to the closest particle, on average, is that to the nearest-neighbor cell.
      For NICCA-SPCA, well-defined peaks also appear that correspond to other underlying lattice distances defined by the square-lattice pattern.
      This is apparent in the distribution functions for $\alpha = 0.2, 0.4$, shown in Fig.\ \ref{fig:RDFFixedBeta}(a), with peaks at distances of $1.0$, $\sqrt{2}\approx{}1.4$, $2.0$, and $\sqrt{5}\approx{}2.2$, etc.
      Increasing the value of $\alpha$ in the NICCA-SPCA regime increases the uncertainty in the position of the particle within the cell, i.e., it leads to peak broadening.

      Now, in the NICCA-MPCA case, the position of the first one or more peaks will depend on the value of $\alpha$ for a given value of $\beta=1.2>1$; also shown in Fig.\ \ref{fig:RDFFixedBeta}(a).
      Additional peaks will reflect the intra-cell particle distribution and can be positioned well below the unit-cell size.
      Peak broadening and peak-peak overlap are superimposed in this case, but the pattern-induced tendency for semi-ordering is still quite visible in the appropriate curves for $\alpha = 0.8, 1.4$ in Fig.~\ref{fig:RDFFixedBeta}(a).

      In the ICCA case, the peak structure and the position of the first peak will be more complex due to the possibility of overlap of an incoming particle with others belonging to neighboring cells.
      In terms of the distribution function, the presence of the excluded volume interaction has an interesting effect of smoothing out the lattice-induced tendency for ordering, as shown in Fig.~\ref{fig:RDFFixedBeta}(b).
      The effect is particularly noticeable when one compares these plots with those for the same values of $\alpha$ in Fig.\ \ref{fig:RDFFixedBeta}(a).
      The features of the distribution function seem to be determined primarily by the particle-particle jamming effects, rather than by the underlying landing-cell pattern.

      Finally, for $\beta= 0.8$, though the interaction between particles at different cells is present, it is not as prominent as for $\beta= 0.2$, and one observes an intermediate behavior, as shown in Fig.\ \ref{fig:RDFFixedBeta}(c).
      The distribution function is still smoothed out due to the jamming effects between particles at different cells, but traces of the lattice-induced ordering remain, especially for $\alpha= 0.2, 0.4$.

    \subsection{Effect of varying $\beta$ on the distribution function}

\begin{figure}[t]
        \includegraphics[width=8.5cm]{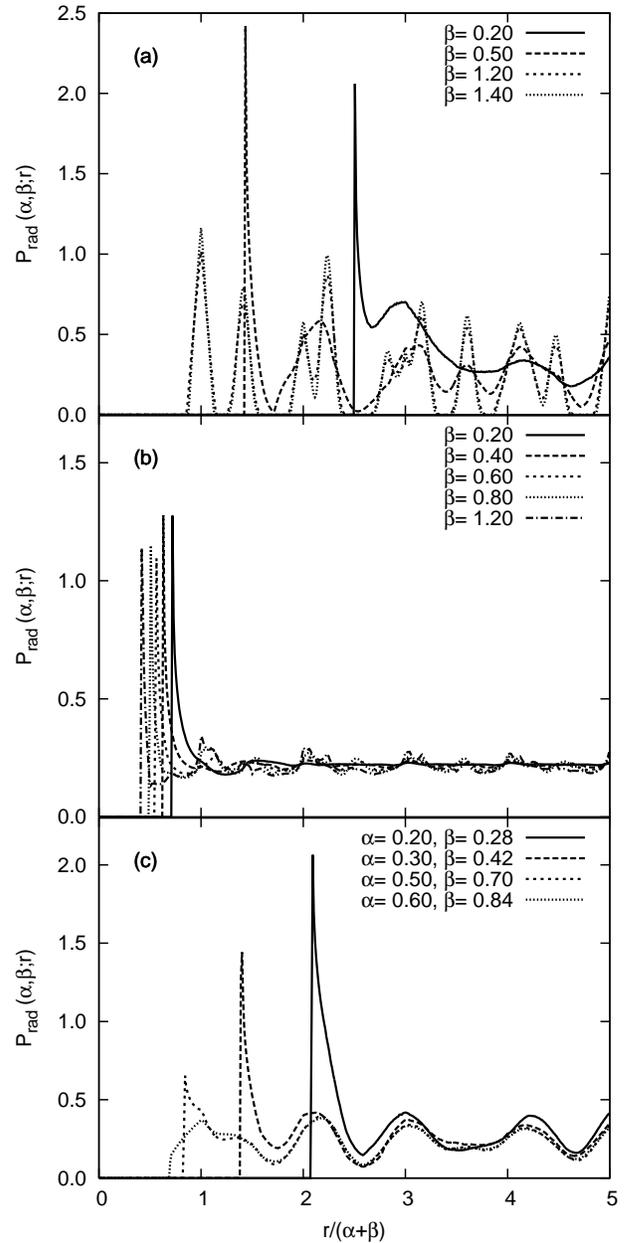}
        \caption
        {The  distribution function
         (a) for constant $\alpha= 0.2$, with $\beta = 0.2$, $0.5$, $1.2$;
         (b) for constant $\alpha= 1.2$, with $\beta = 0.2$, $0.4$, $0.6$, $0.8$, $1.2$;
         (c) along the diagonal line $(0,0)\mbox{---}(1/\sqrt{2},1)$ in the ICCA-SPCA region of the $(\alpha,\beta)$ ``phase diagram,'' Fig.\ \ref{fig:phase.diagram}, with $(\alpha,\beta) = (0.2, 0.28)$, $(0.3, 0.42)$, $(0.5, 0.70)$, $(0.6, 0.84)$.
         \label{fig:RDFBetaDiagonal}
        }
      \end{figure}

      In order to discuss the effect of the value of $\beta$ on the structure of the jammed state, let us first consider fixed $\alpha= 0.2$, with varying $\beta = 0.2$, $0.5$, and $1.2$, as shown in Fig.\ \ref{fig:RDFBetaDiagonal}(a).
      One observes that, as $\beta$ increases, the distribution function becomes more detailed with peaks becoming sharper.
      There is also peak splitting, related to a lesser degree of excluded volume interaction between a particle attempting adsorption and another one from a different cell.
      We comment that for values of $\beta\ge{}1$ the general shape of the radial distribution function is no longer changing, since particles cannot overlap.

      % #############################################################################
      % #############################################################################

      The excluded volume interaction with particles belonging to neighboring cells also reduces the number of particles effectively adsorbed in a cell as observed in snapshots of the jammed state in Figs.\ \ref{fig:snapshotNICCAMPCA} and \ref{fig:snapshotICCAMPCA}, with $\alpha$ fixed at $1.2$, while $\beta$ changing from $1.2$ to $0.2$, respectively.
      For $\alpha= 1.2$ and for increasing values of $\beta$ in the interval $(0,1)$, one observes, see Fig.\ \ref{fig:RDFBetaDiagonal}(b), that the radial distribution function reflects more structure from particle arrangements inside the cells.
      As $\beta$ increases, the position of the first peak also shifts to lower values.
      However, as far as the pattern-induced ordering is concerned, very little trace is left of it, and the curves are relatively flat, dominated by particle-particle jamming effects.

    \subsection{Jammed-state structure along $\beta = \sqrt{2} \alpha$}

      As our last example, we consider the effect of changing both $\alpha$ and $\beta$ along the diagonal of the ICCA-SPCA region, which corresponds to values of $0< \alpha<1/\sqrt{2}$ and $0<\beta<1$.
      Here particles are larger than the cells and they can overlap particles from neighboring cells, not necessarily the nearest-neighbor ones.
      In the ICCA-SPCA regime, particle deposition leads to highly correlated jammed structures.
      Though not specifically taken along the diagonal line, a snapshot of such a highly-correlated jammed state can be seen in Fig.\ \ref{fig:snapshotICCASPCA}.
      As a rule, the tails of the distribution functions, see Fig.\ \ref{fig:RDFBetaDiagonal}(c), mostly coincide regardless of the values of both $\alpha$ and $\beta$.
      This simply reflects the fact that these correlations are dominated by the excluded-volume jamming effects, rather than by the landing-cell pattern.
      However, correlations up to (scaled) distances of $\sim 2.5$ units do show parameter-dependent features.
      Specifically, the position of the first peak shifts to lower values of the distance, since the relative size of the particles compared to that of the unit cells decreases for increasing values of $\alpha$ (or $\beta$).\vspace*{-0.25cm}

    \subsection{Concluding remarks}

      We studied RSA of disk-shaped particles on patterned substrates, specifically, square landing-cells positioned in a square lattice array, with focus on the jammed state properties.
      An efficient numerical algorithm was implemented to simulate the two-dimensional disk deposition.
      The interplay of the two geometrical parameters, the cell size and cell-cell separation, was found to have a striking influence on the deposit morphology and density, as well as on the in-cell particle population.
      We found that the deposit morphologies could be lattice like, locally homogeneous, and locally ordered.
      The distribution function describes the degree to which the cell pattern affects the deposit morphology for various values of the cell size and cell-cell separation parameters.
      This ordering effect competes with the buildup of more random correlations due to the excluded volume interaction.

      Due to the increased use of particles of sizes well below $1\ \!\mu m$, future applications of deposition on patterned substrates will involve features at the nanoscale, which are quite difficult to manufacture with geometrical precision and uniformity. Therefore, possible directions for future work should include consideration of the effects of randomness in the pattern, of particle diffusion within the cells, of particle detachment, as well as of differences in the interactions of the arriving particles with surface features, such as cell interior vs.\ edges.\vspace*{-0.25cm}

  \begin{acknowledgments}
    This research has been funded by a Funda\c c\~ao para a Ci\^encia e a Tecnologia research grant and SeARCH (Services and Advanced Research Computing with HTC/HPC clusters) (under contract CONC-REEQ/443/2001).
    One of us, NA, thanks Funda\c c\~ao para a Ci\^encia e a Tecnologia for a Ph.D.\ fellowship (SFRH/BD/17467/2004).
    AC wants also to thank both Funda\c c\~ao para a Ci\^encia e a Tecnologia and Funda\c c\~ao Calouste Gulbenkian for fellowships to visit Los Alamos National Laboratory.
    AC also acknowledges the warm hospitality of the T-12 Group at Los Alamos National Laboratory.
    VP acknowledges funding of this research by the US National Science Foundation under grant DMR-0509104.
  \end{acknowledgments}

 {\frenchspacing
  \bibliography{Cadilhe}
 }
\end{document}